\documentclass{cidr-2019}

\usepackage{amsmath}
\usepackage{epsfig}
\usepackage{graphicx}
\usepackage{placeins}
\usepackage{url}
\usepackage{subcaption}
\usepackage{paralist}
\usepackage{color}
\usepackage{caption}
\usepackage{multirow}
\usepackage{enumitem}
\usepackage{comment}
\usepackage{listings}
\usepackage{graphicx}
\usepackage{balance}

\makeatletter
\newenvironment{sql}%
 {\vskip 5pt\begin{list}{}{%
  \setlength{\topsep}{0pt}\setlength{\partopsep}{0pt}\setlength{\parskip}{0pt}%
  \setlength{\parsep}{0pt}\setlength{\labelwidth}{0pt}%
  \setlength{\rightmargin}{0pt}\setlength{\leftmargin}{0pt}%
  \setlength{\labelsep}{0pt}%
  \obeylines\@vobeyspaces\normalfont\ttfamily%
  \item[]}}
 {\end{list}\vskip5pt\noindent}
\makeatother

\begin{document}

\title{Exact Selectivity Computation for Modern In-Memory Database Query Optimization}

\numberofauthors{3}

\author{
\alignauthor
Jun Hyung Shin\\
\affaddr{University of California Merced}\\
\email{jshin33@ucmerced.edu}
\alignauthor
Florin Rusu\\
\affaddr{University of California Merced}\\
\email{frusu@ucmerced.edu}
\alignauthor
Alex \c{S}uhan\\
\affaddr{MapD Technologies, Inc.}\\
\email{alex@mapd.com}
}

\maketitle

\begin{abstract}
Selectivity estimation remains a critical task in query optimization even after decades of research and industrial development. Optimizers rely on accurate selectivities when generating execution plans. They maintain a large range of statistical synopses for efficiently estimating selectivities. Nonetheless, small errors -- propagated exponentially -- can lead to severely sub-optimal plans---especially, for complex predicates. Database systems for modern computing architectures rely on extensive in-memory processing supported by massive multithread parallelism and vectorized instructions. However, they maintain the same synopses approach to query optimization as traditional disk-based databases. We introduce a novel query optimization paradigm for in-memory and GPU-accelerated databases based on \textit{exact selectivity computation (ESC)}. The central idea in ESC is to compute selectivities exactly through queries during query optimization. In order to make the process efficient, we propose several optimizations targeting the selection and materialization of tables and predicates to which ESC is applied. We implement ESC in the MapD open-source database system. Experiments on the TPC-H and SSB benchmarks show that ESC records constant and less than 30 milliseconds overhead when running on GPU and generates improved query execution plans that are as much as 32X faster.
\end{abstract}

\section{Introduction}\label{intro}
\begin{figure}[htbp]
  \centering
  \begin{subfigure}{.85\columnwidth}
    \includegraphics[width=\columnwidth]{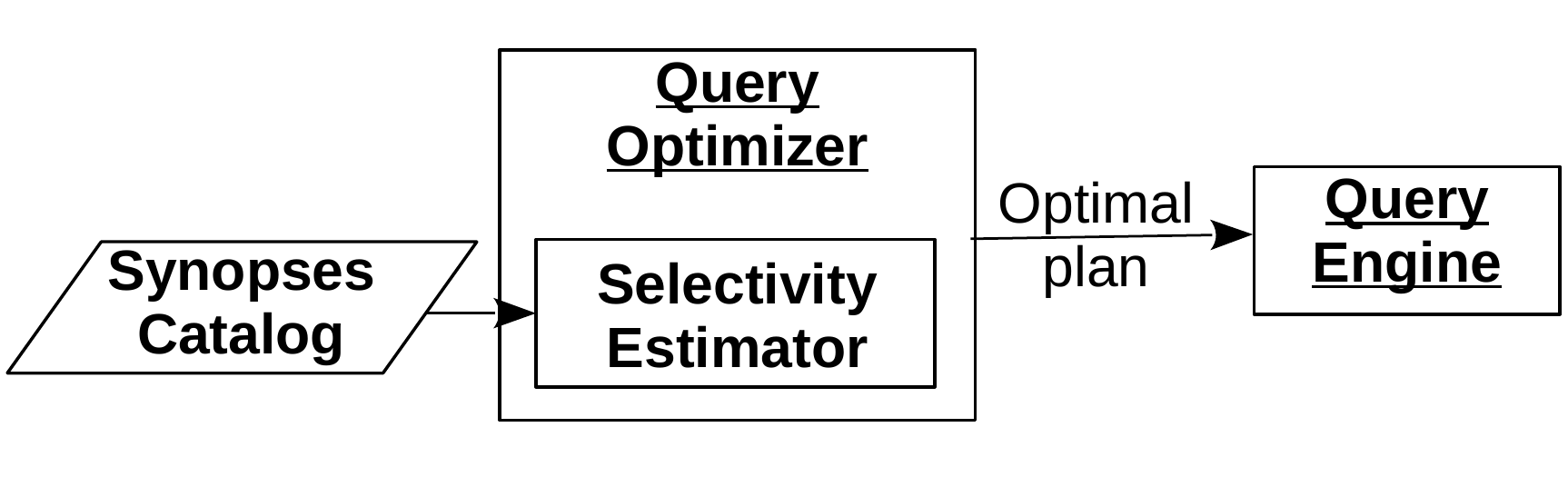}
    \caption{Synopses-driven selectivity estimation}
	  \label{fig:old}
  \end{subfigure}
  \par\medskip
  \begin{subfigure}{.85\columnwidth}
    \includegraphics[width=\columnwidth]{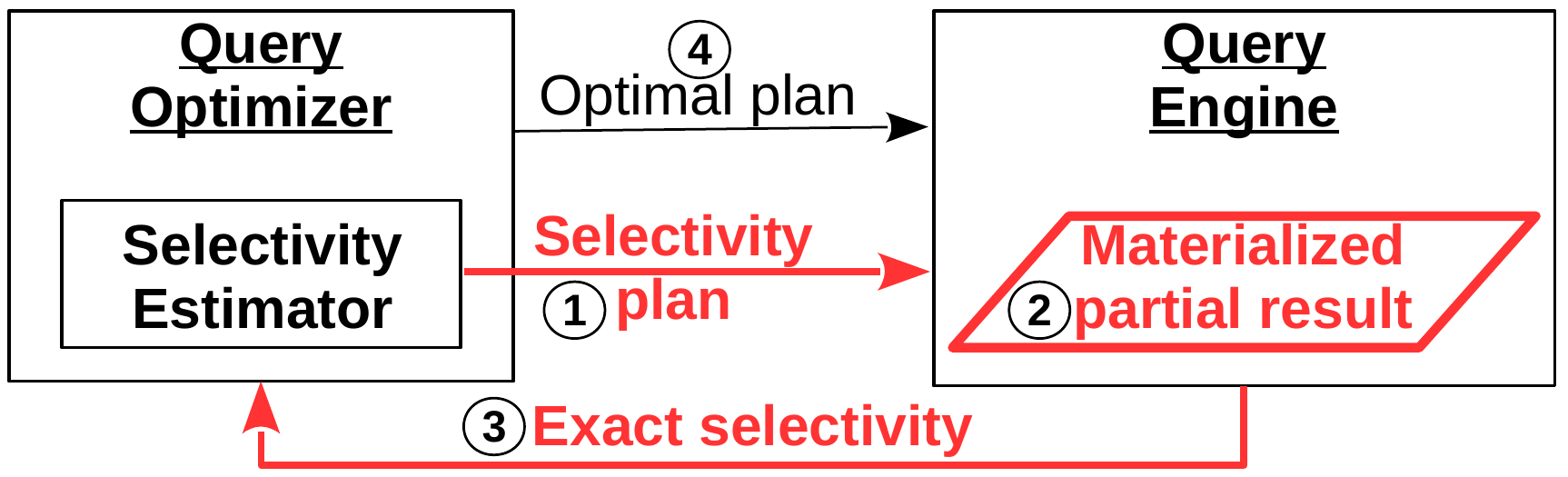}
    \caption{Exact selectivity computation (ESC)}
	  \label{fig:new}
  \end{subfigure}
  \caption{Query optimization strategies.}
\end{figure}

Consider the following \texttt{SQL} query:
\begin{sql}
SELECT R.A, S.B, R.C, R.D
FROM R, S, T
WHERE R.A = S.A AND S.B = T.B AND
\hspace*{0.5cm}R.B = x AND R.C BETWEEN (y1, y2) AND 
\hspace*{0.5cm}(R.D = z1 OR R.D > z2) AND udf(R.B,R.D) > w
\end{sql}\label{query}
The tuples from table \texttt{R} that participate in the join are selected by a complex predicate $\sigma_{B,C,D}(R)$, over three attributes with exact, range, and \texttt{OR} conditions, and a user-defined function \texttt{udf}. When computing the optimal execution plan, i.e., join ordering, the query optimizer has to estimate the selectivity of $\sigma_{B,C,D}(R)$. When available, this is done with precomputed synopses~\cite{Cormode:2012:SMD:2344400.2344401}, e.g., histograms, samples, sketches, stored in the metadata catalog (Figure~\ref{fig:old}). Otherwise, an arbitrary guess is used, e.g., for \texttt{udf}. Synopses are typically built for a single attribute and assume uniformity and/or independence when they are combined across multiple attributes. These are likely to miss correlations between attributes over skewed or sparse data and result in inaccurate estimates which produce highly sub-optimal query execution plans~\cite{Leis:good-optimizer}. We argue that -- despite the large volume of work on the topic~\cite{Chaudhuri:1998:OQO:275487.275492,Moerkotte:optim-bound,Tzoumas:optim-graphical} -- selectivity estimation is still an open problem~\cite{lohman:sigmod-blog}.

Database systems for modern computing architectures rely on extensive in-memory processing supported by massive multithread parallelism and vectorized instructions. GPUs represent the pinnacle of such architectures, harboring thousands of SMT threads which execute tens of vectorized SIMD instructions simultaneously. MapD\footnote{\url{https://www.mapd.com/}}, Ocelot\footnote{\url{https://bitbucket.org/msaecker/monetdb-opencl}}, and CoGaDB\footnote{\url{http://cogadb.dfki.de/}} are a few examples of modern in-memory databases with GPU support. They provide relational algebra operators and pipelines for GPU architectures~\cite{He:2009:RQC:1620585.1620588,Bress:2014,Funke:2018:PQP:3183713.3183734} that optimize memory access and bandwidth. However, they maintain the same synopses approach to query optimization as traditional disk-based databases.

We introduce a novel query optimization paradigm for in-memory and GPU-accelerated databases based on \textit{exact selectivity computation (ESC)}. The central idea in ESC is to compute selectivities exactly through queries during query optimization. As illustrated in Figure~\ref{fig:new}, our approach interacts with the query execution engine, while the synopses-driven solution only accesses the system catalog. For the given query example, the optimizer instructs the execution engine to first perform the selectivity sub-query:
\begin{sql}
SELECT R.A, R.C, R.D
FROM R
WHERE R.B = x AND R.C BETWEEN (y1, y2) AND 
\hspace*{0.5cm}(R.D = z1 OR R.D > z2) AND udf(R.B,R.D) > w
\end{sql}\label{query:sel-comp}
in order to compute the cardinality of $\sigma_{B,C,D}(R)$ exactly. This value is used together with the cardinalities of \texttt{S} and \texttt{T} to compute the best join order in the optimal query plan. Moreover, the result of the selectivity sub-query is temporarily materialized and reused instead of \texttt{R} in the optimal execution plan. This avoids re-executing the selection $\sigma_{B,C,D}(R)$ during the complete query execution. In order to make the process efficient, we propose optimizations targeting the selection of tables and predicates to which ESC is applied.

While it is clear that the plan computed by ESC is better -- or at least as good -- the impact on query execution time depends on the ratio between the execution time for the selectivity sub-query and the original plan. The assumption we make is that the sub-query execution is relatively negligible---valid for in-memory databases. We have to consider two cases. First, if the new query plan is improved by a larger margin than the sub-query time, the total execution time is reduced. We argue that exact selectivities are likely to achieve this for queries over many tables and with complex predicates. In the second case, the optimal query plan, i.e., join order, does not change even when selectivities are exact. Materialization minimizes the overhead incurred by the sub-query through subsequent reuse further up in the plan. In-memory databases prefer materialization over pipelining~\cite{Mostak:2013,Funke:2018:PQP:3183713.3183734} due to better cache access.

Our contributions are summarized as follows:
\begin{itemize}
\item We introduce ESC as a novel paradigm to query optimization that is applicable independent of the predicate complexity and data distribution. ESC does not rely on any database synopses. It can be applied to any modern in-memory and GPU-accelerated database.
\item We implement ESC in the open-source MapD database with minimal changes to the query optimizer. The code is adopted by the MapD core engine.
\item We perform extensive experiments over the TPC-H and Star-Schema (SSB) benchmarks. The results show that ESC records constant and less than 30 milliseconds overhead when running on GPU and generates improved query execution plans that are as much as 32X faster for queries with complex predicates.
\end{itemize}

\section{Exact Selectivity Computation}\label{new_approach}
\begin{figure}[htbp]
  \centering
  \includegraphics[width=\columnwidth]{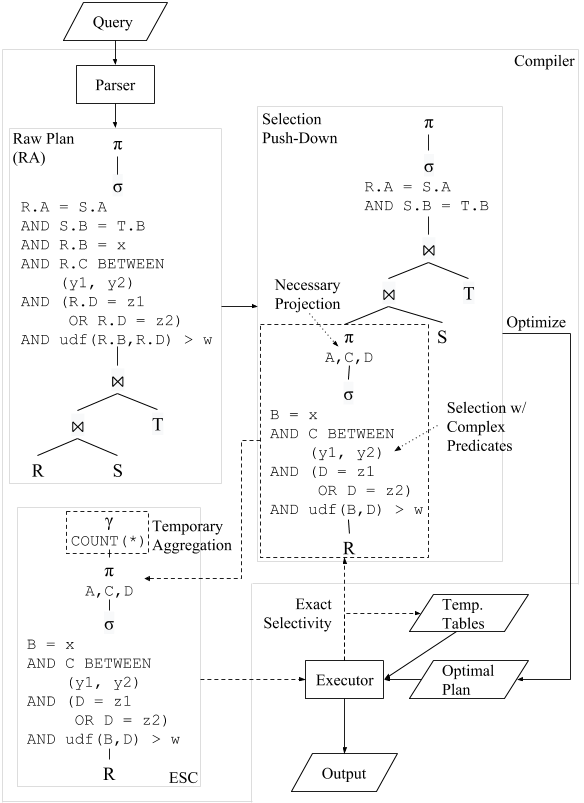}
  \caption{Exact selectivity computation (ESC) workflow.}
  \label{fig:flowchart}
\end{figure}

Figure~\ref{fig:flowchart} depicts the exact selectivity computation workflow corresponding to the example query given in the introduction. Once the query reaches the query compiler, it is first parsed into an abstract syntax tree representation. The compiler searches this representation for selection predicates that can be pushed-down to reduce the cost of subsequent join operations. In the example query, the only selection predicate is $\sigma_{B,C,D}(R)$ on table \texttt{R}. The query optimizer has to determine whether the identified selection(s) should be pushed-down or not based on their cardinality. We assume a materialized query execution strategy since predicates are always pushed-down with pipelining. At this point, the cardinality of $\sigma_{B,C,D}(R)$ has to be estimated to determine the optimal plan. However, this is rather problematic in the case of the complex conditions in $\sigma_{B,C,D}(R)$ and errors may lead to a sub-optimal plan. This is because all the existing selectivity estimation methods either make simplifying assumptions about the distribution of an attribute and the correlations across attributes~\cite{Chaudhuri:1998:OQO:275487.275492}, or -- when such assumptions are not made -- they are applicable to a limited class of (conjunctive) predicates~\cite{Tzoumas:optim-graphical}.

The idea in ESC is to focus entirely on the accuracy of the predicate cardinality. Thus, we compute this value exactly with a simple aggregate query:
\begin{sql}
SELECT COUNT(*)
FROM R
WHERE R.B = x AND R.C BETWEEN (y1, y2) AND 
\hspace*{0.5cm}(R.D = z1 OR R.D > z2) AND udf(R.B,R.D) > w
\end{sql}\label{query:sel-comp-agg}
There are several steps to generate the exact selectivity computation query from the original relational algebra tree (RA). First, the compiler pushes-down an additional projection with the attributes required by subsequent operators in the plan---in this case, \texttt{A}, \texttt{C}, and \texttt{D}. Although this is not necessary for the selectivity computation itself, it reduces data access in column-wise in-memory databases. If no column is included in the projection, the execution engine accesses unnecessary columns, e.g., \texttt{B}, which incurs overhead. Hence, the compiler has to find all the columns required for further processing and push them all down. At this point, we have a simple \texttt{SELECT-FROM-WHERE} sub-tree corresponding to the selectivity sub-query. To compute its cardinality, the compiler adds a temporary aggregate \texttt{COUNT} operator, as shown at the bottom-left of Figure~\ref{fig:flowchart}. The generated tree is then passed to the query engine for execution. Upon completion, the exact cardinality is returned to the query optimizer instead of an estimate computed from synopses stored in the catalog. The optimizer uses the computed cardinality to determine whether pushing-down the selection is effective or not. Since intermediate results are materialized, this depends on the selectivity of $\sigma_{B,C,D}(R)$. If the selection is pushed-down, the result of the selectivity sub-query is materialized at optimization. Otherwise, it is discarded entirely. The same procedure is applied to each table having selection predicates. After all the selectivities are computed, the query optimizer proceeds to computing join ordering in the optimal plan.

\begin{figure}[htbp]
  \centering
  \includegraphics[width=\columnwidth]{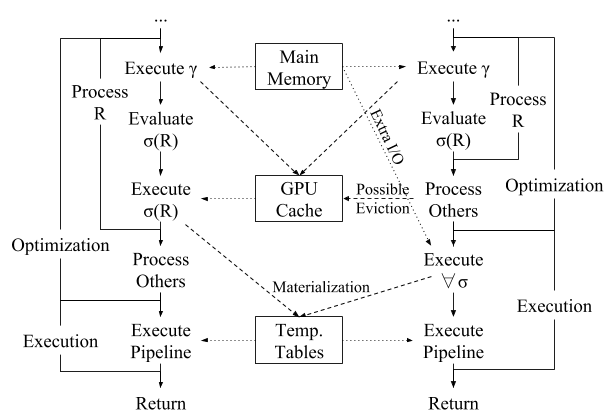}
  \caption{Materialization of $\sigma_{B,C,D}(R)$ during optimization (left) vs. execution (right).}
  \label{fig:when_to_materialize}
\end{figure}

\subsection{Selection Materialization}\label{ssec:esc:materialization}

Exact selectivity computation with selection push-down introduces two problems for materialized execution in memory databases: \textit{When to materialize?} and \textit{Where to materialize?} The materialization incurred by selection push-down is useful only when a significant number of tuples are pruned-out. Otherwise, the function call overhead coupled with the additional memory read/write access do not provide a significant reduction over the non-optimized plan. Moreover, materialization makes sense only for large tables which have significant margin for cardinality reduction. Our initial solution allows the user to specify two parameters that control materialization: the minimum table size and the maximum selectivity threshold. ESC is applied only to tables having more tuples than the minimum size parameter. If this condition is satisfied, \textit{materialization is performed only when the predicate selectivity is below the maximum threshold}. Setting the value of these parameters has an important impact on performance. Currently, this is done by the DBA, however, in the future, we plan to devise adaptive algorithms that determine the optimal values automatically.

If a selection is pushed-down, it eventually generates an intermediate result that is materialized at some point during execution. In ESC, as shown in Figure~\ref{fig:when_to_materialize}, \textit{materialization is performed by the query optimizer} instead of the execution engine. This facilitates the immediate reuse of the selectivity sub-query output which is already cached in (GPU) memory. If we delay materialization to execution, we cannot guarantee that the sub-query output is still cached, thus additional memory access being necessary. The result set is materialized as a temporary table.

\begin{figure}[htbp]
  \centering
  \begin{subfigure}{0.49\columnwidth}
    \includegraphics[width=\linewidth]{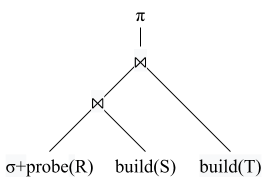}
    \caption{Left-deep plan}
    \label{fig:plan_naive}
  \end{subfigure}
  \par
  \begin{subfigure}{0.49\columnwidth}
    \includegraphics[width=\linewidth]{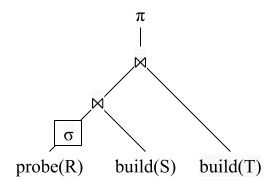}
    \caption{ESC on the probe input}
    \label{fig:plan_probe}    
  \end{subfigure}
  \begin{subfigure}{0.49\columnwidth}
    \includegraphics[width=\linewidth]{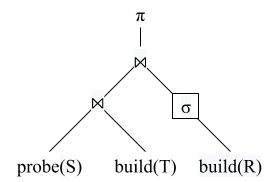}
    \caption{ESC on the build input}
    \label{fig:plan_build}    
  \end{subfigure}
  \caption{Interaction between ESC and join ordering.}
  \label{fig:plans}
\end{figure}

\subsection{Join Ordering}\label{ssec:esc:join_ordering}

Based on the computed exact selectivities, the optimizer finds an optimal join order that minimizes the intermediate join output sizes. We start from a typical left-deep tree (Figure~\ref{fig:plan_naive}) in which all the joins are implemented with hashing. Hash tables are built for all the relations except the largest, which is used for probing the built hash tables. The reason we probe on the largest relation is limited memory capacity---only a few GB on GPU. Even if there is sufficient memory, building a hash table for the largest relation is more expensive. First, we join the smallest relation that shares a join predicate with the probed input in order to avoid expensive Cartesian products, and so on. All selection predicates are evaluated during probing since no selection is pushed down. Code generation techniques are applied to combine all the selections and probings in a single operator that does not materialize any intermediate results during execution.

We consider the effect of materialization from each exact selectivity computation and decide whether to push-down the selection or not. A selection push-down clearly reduces the number of tuples to be joined, however, it also adds materialization which incurs extra read/write access and can result in performance degradation overall. We have two cases to consider---selection push-down on the probing relation and on the building relation, respectively. In the former case, as depicted in Figure~\ref{fig:plan_probe}, selection push-down is not beneficial because it splits probing into two separate condition evaluations with a materialization in-between. This impacts negatively the memory access. Therefore, we do not perform selection push-down on the (largest) probing relation. In the latter case, as depicted in Figure~\ref{fig:plan_build}, selections are pushed-down since materialization reduces their size and this can result in an improved join ordering. This happens only for selective predicates, though. Selections with high selectivity, however, are discarded when the cost to process the materialization is more expensive than the benefit of selection push-down. The optimal execution plan pushes-down only those selections that overcome the materialization overhead.

\subsection{Advantages and Limitations}\label{ssec:esc:analysis}

The ESC strategy has several strengths over the synopses-driven approach:
\begin{itemize}[leftmargin=*,noitemsep,nolistsep]
\item \textit{Exact}: Needless to say, the core of ESC stems from computing the exact selectivity by running an additional aggregate query on demand. ESC is no longer a cardinality estimation method; it is an exact operation. Thus, its accuracy is no longer affected by the number of attributes or their distribution. Moreover, the computed selectivity is also consistent no matter how frequent the database is updated---no maintenance is required.
\item \textit{Comprehensible}: ESC is rather simple since it does not require any complicate synopses or statistical comprehension for estimation. All we need is building an aggregate query and its execution to retrieve the exact selectivity.
\item \textit{Easy to integrate}: From the perspective of an actual implementation, ESC requires only a handful of modifications to the query optimizer in order to compute the exact selectivity. The execution engine is kept unchanged.
\end{itemize}

Nevertheless, the time to execute a query to retrieve the exact selectivity is larger than synopses-driven estimation. If the total overhead is larger than the improvement gained from a better query execution plan, ESC does not provide any benefit. Given that the overhead is mostly due to I/O access, ESC is highly-dependent on the underlying database system architecture. For example, ESC does not work well on a disk-based database. Modern in-memory and GPU-accelerated databases, however, have a much lower overhead for executing simple aggregate queries.

\subsection{Related Work}\label{ssec:esc:rel-work}

\textbf{Synopses-driven selectivity estimation.}
A large variety of synopses have been proposed for selectivity estimation. Many of them are also used in approximate query processing~\cite{Cormode:2012:SMD:2344400.2344401}. Histograms are the most frequently implemented in practice. While they provide accurate estimation for a single attribute, they cannot handle correlations between attributes as the dimensionality increases~\cite{Poosala:1997:SEW:645923.673638}. The same holds for sketches~\cite{Rusu:2007:SAS:1247480.1247503}. Samples are sensitive to skewed and sparse data distributions~\cite{Wu:2012}. Although graphical models~\cite{Tzoumas:optim-graphical} avoid all these problems, they are limited to conjunctive queries and perform best for equality predicates. Since ESC is exact, it achieves both accuracy and generality.

\textbf{Query re-optimization.}
The LEO optimizer~\cite{Markl:2003:LAQ:1014767.1014781} monitors relational operator selectivities during query execution. These exact values are continuously compared with the synopses-based estimates used during optimization. When the two have a large deviation, synopses are adjusted to automatically correct their error for future queries. Mid-query re-optimization~\cite{Kabra:1998:EMR:276304.276315} takes immediate action and updates the execution plan for the running query. This, however, requires dropping already computed partial results which introduces considerable overhead. ESC computes the exact cardinalities at optimization and applies them to the current query which has the best plan from the beginning.

\section{Experimental Evaluation}\label{experiments}
The purpose of the experimental evaluation is to quantify the impact ESC has on query execution performance in modern in-memory and GPU-accelerated databases. Specifically, we measure the time overhead incurred by the selectivity sub-query and the overall query execution time with and without ESC. To this end, we implement ESC in the query optimizer of open-source \texttt{MapD}~\cite{Mostak:2013}---a highly-parallel database engine with extensive GPU support. We name our implementation \texttt{MapD+ESC}. MapD has a materialized query execution strategy which does not perform selection push-down in order to minimize memory access. Moreover, query compilation is applied to fuse non-blocking operators together. The MapD optimizer does not perform any selectivity estimation. Thus, the ESC extension is entire overhead. ESC does not modify the join cardinality estimation which is standard in MapD.

We perform the experiments on a dual-socket server with two 14-core Intel Xeon E5-2660 v4 CPUs (56 threads overall, 256 GB memory) and an NVIDIA Tesla K80 GPU with 2496 cores, a 32-wide SIMD unit, and 24 GB memory. The server runs Ubuntu 16.04 SMP with CUDA 9.1. All the experiments are performed both on CPU and GPU. Switching between the two implementations is done with a command-line parameter in MapD. The results are collected by running each query at least 5 times and reporting the median value. Caches are always hot. Time is measured with the MapD built-in function.

\begin{figure*}[htbp]
  \centering
    \begin{subfigure}{0.49\textwidth}
      \includegraphics[width=\linewidth]{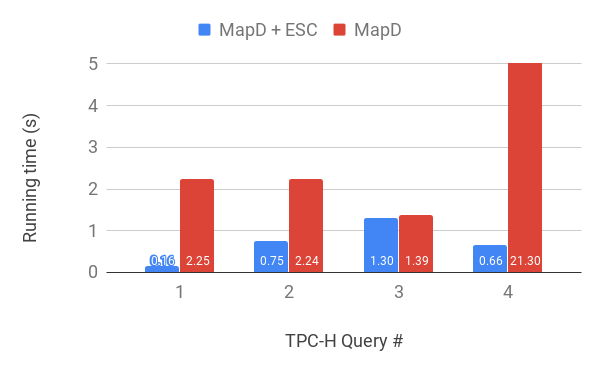}
      \caption{GPU}
      \label{fig:tpch-gpu}    
    \end{subfigure}
    \begin{subfigure}{0.49\textwidth}
      \includegraphics[width=\linewidth]{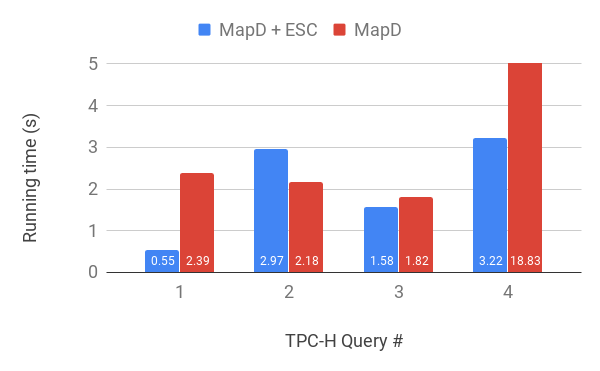}
      \caption{CPU}
      \label{fig:tpch-cpu}    
    \end{subfigure}
    \caption{Execution time (seconds) for the TPC-H (scale 50) queries used in~\cite{Tzoumas:optim-graphical}.}
    \label{fig:eg_tpch}
\end{figure*}

\subsection{ESC Overhead}\label{ssec:experiments:overhead}

In order to determine if ESC is feasible in practice, we measure the overhead introduced by exact selectivity computation as a function of the data size, the selectivity of the predicate, and the number of attributes in the selection. We report the execution time of the selectivity sub-query(ies). This means that MapD+ESC executes an additional sub-query for each table having selection predicates. We deactivate push-down materialization to guarantee that MapD+ESC uses exactly the same execution plan as MapD. This allows us to isolate the overhead incurred by ESC. All the experiments in this section are performed on TPC-H data at different scales.

\textbf{Overhead by scale factor.}
We execute the following query template:
\begin{sql}
SELECT COUNT(*)
FROM lineitem, [orders O] [part P] [supplier S]
[WHERE l\_orderkey=o\_orderkey AND o\_orderkey=x]
[WHERE l\_partkey=p\_partkey AND p\_partkey=y]
[WHERE l\_suppkey=s\_suppkey AND s\_suppkey=z]
\end{sql}
which joins \texttt{lineitem} with one of the \texttt{orders}, \texttt{part}, or \texttt{supplier} tables. An equality selection predicate with a constant x, y, or z is applied to the variable table. We run these queries at scale factors 1, 10, and 50. Thus, MapD+ESC performs an additional selectivity sub-query on tables having an increasing number of tuples. Table~\ref{table:overhead_sf} contains the results. We observe that the overhead incurred by MapD+ESC is at most 70 ms. On GPU, the overhead is always below 30 ms. This is a very small price for getting the exact selectivity. When the data size increases, there is almost no difference on the GPU, while on the CPU there is only a factor of 3.5 increase---compared to 50 on data size. This shows that the GPU throughput is much larger due to its considerably higher parallelism. The reason is the different memory hierarchy across the two architectures---the cache overhead increases with data size on CPU.

\begin{table}[htbp]
  \centering
  \caption{Overhead by scale factor (ms).}
  \label{table:overhead_sf}
  \begin{tabular}{|l||*{9}{c|}}
  	\hline
    scale & \multicolumn{3}{c|}{1} & \multicolumn{3}{c|}{10} & \multicolumn{3}{c|}{50}\\
    \hline
    table & O & P & S & O & P & S & O & P & S\\
		\hline
		\hline
		GPU & 21 & 22 & 22 & 20 & 23 & 28 & 21 & 21 & 19\\
    CPU & 21 & 24 & 21 & 49 & 25 & 21 & 70 & 66 & 57\\
    \hline
  \end{tabular}
\end{table}

\textbf{Overhead by selectivity.}
We run the following query:
\begin{sql}
SELECT COUNT(*)
FROM lineitem, orders
WHERE l\_orderkey=o\_orderkey AND o\_orderkey<u 
\end{sql}
in order to measure the impact of predicate selectivity on overhead. u is a parameter that controls the number of tuples from \texttt{orders} participating in the join. u takes values such that the selectivity varies from 0.001\% to 100\%. These correspond to a range from 7.5K to 7.5M on TPC-H scale 50. The results are included in Table~\ref{table:overhead_sel}. As expected, we observe that the overhead increases with selectivity. However, the increase is almost negligible. This is because the selectivity sub-query is not impacted by the predicate selectivity---it is a simple aggregate. Selectivity changes only the number of tuples used to build the hash table on \texttt{orders}. Since MapD+ESC preserves the same plan, the overhead is due entirely to the additional sub-query---similar to Table~\ref{table:overhead_sf}.

\begin{table}[htbp]
  \centering
  \caption{Overhead by selectivity (ms).}
  \label{table:overhead_sel}
  \begin{tabular}{|l||*{7}{r|}}
  	\hline
    selectivity (\%) & 0.001 & 0.01 & 0.1 & 1 & 10 & 100\\
    \hline
    \hline
    GPU & 18 & 24 & 20 & 24 & 24 & 30\\
    CPU & 69 & 73 & 76 & 71 & 72 & 75\\
    \hline
  \end{tabular}
\end{table}

\begin{figure*}[htbp]
  \centering
  \begin{subfigure}{0.49\textwidth}
    \includegraphics[width=\linewidth]{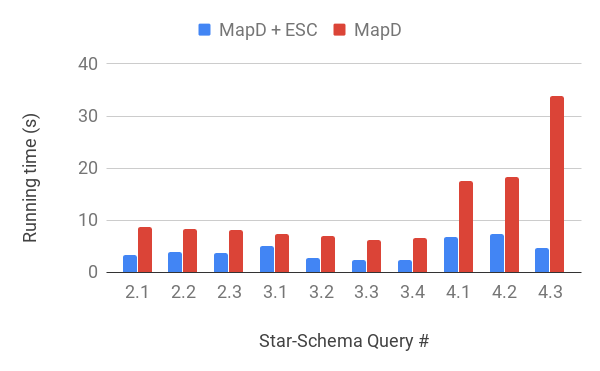}
    \caption{GPU}
    \label{fig:ssb-gpu}    
  \end{subfigure}
  \begin{subfigure}{0.49\textwidth}
    \includegraphics[width=\linewidth]{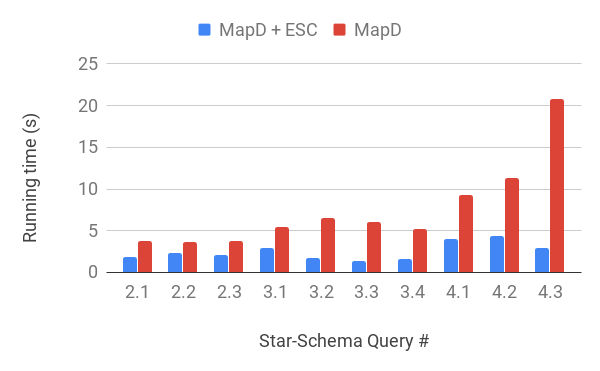}
    \caption{CPU}
    \label{fig:ssb-cpu}    
  \end{subfigure}
  \caption{Execution time (seconds) for the SSB (scale 80) queries.}
  \label{fig:eg_ssb}
\end{figure*}

\textbf{Overhead by number of attributes.}
In order to quantify the impact of the number of attributes and their type on ESC, we run the following set of queries which increase the number of attributes in the predicate progressively:
\begin{sql}
SELECT COUNT(*)
FROM lineitem, orders
WHERE l\_orderkey=o\_orderkey AND o\_orderkey=x
\hspace*{.25cm} [AND o\_custkey=y]
\hspace*{.25cm} [AND o\_orderstatus=z]
\hspace*{.25cm} [AND o\_totalprice=t]
\end{sql}
Moreover, the type of \texttt{o\_orderstatus} and \texttt{o\_totalprice} is \texttt{TEXT} and \texttt{DECIMAL(15,2)}, increasing the evaluation time of the condition compared to \texttt{INTEGER}. All the predicates are equality with constants x, y, z, and t, respectively. We set their values such that the result is always positive. Table~\ref{table:overhead_attr} shows the results for TPC-H scale 50. For 1 attribute, we obtain results that are consistent with Table~\ref{table:overhead_sf}. As the number of attributes increases, the overhead stays constant on GPU because the simple operations in the predicate do not fill the available processing throughput or memory bandwidth. This is not the case on CPU, where we observe a significant increase as we go from 2 to 3 attributes. The main reason is the degradation in cache performance due to a larger number of attributes being accessed. Nonetheless, an overhead of 300 ms may be offset by a better plan computed with exact selectivities.

\begin{table}[htbp]
  \centering
  \caption{Overhead by number of attributes (ms).}
  \label{table:overhead_attr}
  \begin{tabular}{|l||*{4}{r|}}
  	\hline
    \# attributes & 1 & 2 & 3 & 4\\
    \hline
    \hline
    GPU & 21 & 23 & 23 & 18\\
    CPU & 70 & 91 & 273 & 339\\
    \hline
  \end{tabular}
\end{table}

\subsection{Benchmark Queries}\label{ssec:experiments:benchmarks}

We perform an end-to-end comparison between MapD and our extension MapD+ESC on several queries from standard benchmarks. We report the complete query execution time which -- for MapD+ESC -- includes the additional sub-queries and their materialization. In this case, the join order can be different across the two systems since they do not execute the same plan.

\textbf{TPC-H scale 50.}
We execute the four queries introduced in~\cite{Tzoumas:optim-graphical} (Appendix B). Each of them is defined over 3 tables and includes a selection for each table. These queries have selectivities that are difficult to estimate because of correlations across tables. ESC eliminates this issue completely. Figure~\ref{fig:eg_tpch} depicts the results. With a single exception on CPU, MapD+ESC always outperforms standard MapD. The improvement ranges from 1.06X to 32.27X. It is larger when MapD+ESC selects a better execution plan. This is directly correlated with the predicate selectivity on each of the tables. While there is not much difference between MapD running on CPU and GPU, MapD+ESC is always superior on GPU---by as much as 4.88X. This is because the overhead on GPU is much smaller. Moreover, materialization on CPU can interfere with caching and this can have unexpected consequences---the case for query 2.

\textbf{SSB scale 80.}
We execute all the queries from the Star-Schema Benchmark (SSB) that are defined over at least two tables. Unlike TPC-H, the fact table \texttt{lineorder} in SSB is much larger and takes 99\% of the total dataset. Consequently, the dimension tables are relatively small and this can be a problem for ESC. Figure~\ref{fig:eg_ssb} presents the results. MapD+ESC provides faster execution times across all the queries, both on CPU and GPU. As the number of tables and predicates in a query increases, the gap becomes larger. The largest improvement -- more than 7X -- is gained for query 4.3 which is more selective. As before, the main reason for the speedup is the ability of MapD+ESC to identify better query plans because of exact selectivities. Interestingly enough, the CPU implementation is faster than the GPU in this case. The larger scale factor is the main culprit. The amount of data that has to be moved to the GPU saturates the PCIe bus and this increases execution time for all the queries.

\section{Conclusions and Future Work}\label{conclusions}
In this paper, we introduce a novel query optimization paradigm ESC for in-memory and GPU databases. The central idea in ESC is to compute selectivities exactly through queries during query optimization. We propose several optimizations targeting the selection and materialization of tables and predicates. We implement ESC in the MapD open-source database system. Experiments on the TPC-H and SSB benchmarks show that ESC records constant and less than 30 milliseconds overhead when running on GPU and generates improved query execution plans that are as much as 32X faster. ESC improves the execution both on CPU as well as on GPU. In future work, we plan to design adaptive algorithms that can detect automatically when materialization is feasible. We also plan to investigate strategies that split predicate selectivity evaluation.

\small{
\bibliographystyle{abbrv}

\begin{thebibliography}{10}

\bibitem{Bress:2014}
S.~Bre{\ss}, M.~Heimel, N.~Siegmund, L.~Bellatreche, and G.~Saake.
\newblock {GPU-accelerated Database Systems: Survey and Open Challenges}.
\newblock In {\em TLDKS}, pages 1--35. 2014.

\bibitem{Chaudhuri:1998:OQO:275487.275492}
S.~Chaudhuri.
\newblock {An Overview of Query Optimization in Relational Systems}.
\newblock In {\em PODS 1998}.

\bibitem{Cormode:2012:SMD:2344400.2344401}
G.~Cormode, M.~Garofalakis, P.~Haas, and C.~Jermaine.
\newblock {Synopses for Massive Data: Samples, Histograms, Wavelets, Sketches}.
\newblock {\em Found. Trends Databases}, 4(1), 2012.

\bibitem{Funke:2018:PQP:3183713.3183734}
H.~Funke, S.~Bre\ss, S.~Noll, V.~Markl, and J.~Teubner.
\newblock {Pipelined Query Processing in Coprocessor Environments}.
\newblock In {\em SIGMOD 2018}.

\bibitem{He:2009:RQC:1620585.1620588}
B.~He, M.~Lu, K.~Yang, R.~Fang, N.~Govindaraju, Q.~Luo, and P.~Sander.
\newblock {Relational Query Coprocessing on Graphics Processors}.
\newblock {\em ACM TODS}, 34(4), 2009.

\bibitem{Kabra:1998:EMR:276304.276315}
N.~Kabra and D.~DeWitt.
\newblock {Efficient Mid-query Re-optimization of Sub-optimal Query Execution
  Plans}.
\newblock In {\em SIGMOD 1998}.

\bibitem{Leis:good-optimizer}
V.~Leis, A.~Gubichev, A.~Mirchev, P.~Boncz, A.~Kemper, and T.~Neumann.
\newblock {How Good Are Query Optimizers, Really?}
\newblock {\em PVLDB}, 9(3), 2015.

\bibitem{lohman:sigmod-blog}
G.~Lohman.
\newblock {Is Query Optimization a ``Solved'' Problem?}, 2014.
\newblock [Online at: \url{http://wp.sigmod.org/?p=1075}].

\bibitem{Markl:2003:LAQ:1014767.1014781}
V.~Markl, G.~Lohman, and V.~Raman.
\newblock {LEO: An Autonomic Query Optimizer for DB2}.
\newblock {\em IBM Syst. J.}, 42(1), 2003.

\bibitem{Moerkotte:optim-bound}
G.~Moerkotte, T.~Neumann, , and G.~Steidl.
\newblock {Preventing Bad Plans by Bounding the Impact of Cardinality
  Estimation Errors}.
\newblock {\em PVLDB}, 2(1), 2009.

\bibitem{Mostak:2013}
T.~Mostak.
\newblock {An Overview of MapD (Massively Parallel Database)}.
\newblock {\em MIT white paper}, 2013.

\bibitem{Poosala:1997:SEW:645923.673638}
V.~Poosala and Y.~Ioannidis.
\newblock {Selectivity Estimation Without the Attribute Value Independence
  Assumption}.
\newblock In {\em VLDB 1997}.

\bibitem{Rusu:2007:SAS:1247480.1247503}
F.~Rusu and A.~Dobra.
\newblock {Statistical Analysis of Sketch Estimators}.
\newblock In {\em SIGMOD 2007}.

\bibitem{Tzoumas:optim-graphical}
K.~Tzoumas, A.~Deshpande, and C.~S. Jensen.
\newblock {Lightweight Graphical Models for Selectivity Estimation Without
  Independence Assumptions}.
\newblock {\em PVLDB}, 4(11), 2011.

\bibitem{Wu:2012}
W.~Wu.
\newblock {Sampling-Based Cardinality Estimation Algorithms: A Survey and An
  Empirical Evaluation}, 2012.

\end{thebibliography}

}

\end{document}